# Adaptive boundary conditions for exterior stationary flows in three dimensions

Vincent Heuveline

Computing Center and

Institute for Applied Mathematics

University of Karlsruhe

vincent.heuveline@rz.uni-karlsruhe.de

Peter Wittwer*

Département de Physique Théorique

Université de Genève, Switzerland

peter.wittwer@physics.unige.ch

**Abstract**

Recently there has been an increasing interest for a better understanding of ultra low Reynolds number flows. In this context we present a new setup which allows to efficiently solve the stationary incompressible Navier-Stokes equations in an exterior domain in three dimensions numerically. The main point is that the necessity to truncate for numerical purposes the exterior domain to a finite sub-domain leads to the problem of finding so called "artificial boundary conditions" to replace the conditions at infinity. To solve this problem we provide a vector filed that describes the leading asymptotic behavior of the solution at large distances. This vector field depends explicitly on drag and lift which are determined in a self-consistent way as part of the solution process. When compared with other numerical schemes the size of the computational domain that is needed to obtain the hydrodynamic forces with a given precision is drastically reduced, which in turn leads to an overall gain in computational efficiency of typically several orders of magnitude.

**Mathematics Subject Classification (2000).** 76D05, 76D25, 76M10, 41A60, 35Q35.

**Keywords.** Navier-Stokes equations, artificial boundary conditions, drag, lift.

## 1 Introduction

There is an increasing interest in considering applications leading to exterior flow problems at Reynolds numbers of the order of one to several thousand. Specific examples of situations where such flows occur are the sedimentation of small particles in the context of climate prediction [1, 23] and the engineering of so called micro-air vehicles (MAV) [19, 7, 9, 8]. In the first example sedimentation speeds have to be computed accurately and in the second case the prediction of performance requires that the entire flow field be calculable in detail and that the hydrodynamic forces be determined with high precision. Linearized theories (Stokes, Oseen) provide such a quantitative description (better than one percent) for Reynolds numbers less than one [2] and traditional approximation schemes based on some version of boundary layer theory work well for Reynolds numbers exceeding some ten thousand [7]. For the intermediate regime where neither the viscous nor the inertial forces dominate the full Navier-Stokes equations need to be solved [2]. However, when truncating for numerical purposes the infinite exterior domain to a finite sub-domain one is confronted with the problem of finding so called "artificial boundary conditions" on the outer boundary of this sub-domain in order to replace the boundary conditions at infinity [5]. It turns out that for the Reynolds numbers under consideration any naïve choice of such boundary conditions modifies the hydrodynamic forces significantly, unless excessively large computational domains are used.

The scheme of artificial boundary conditions proposed here consists of an explicit expression for a vector field that describes the solution to leading order at large distances from the body. This vector field depends explicitly on the hydrodynamic forces which are determined in a self-consistent way as part of the solution process. This method has been introduced in [5] for the analogous two dimensional problem on the bases of recent analytic work [25, 26]. In [6] the method has been further refined through the use of additional analytical input [13], and the efficiency of the scheme is now widely recognized: in [3] the same ideas have been successfully tested on the semi-infinite flat plate problem and in [18] the scheme has been implemented in the context of lattice Boltzmann simulations. The present generalization of the setup to three dimensions is again based on analytic work [27] and when compared with results obtained using traditional boundary conditions computational times are again typically reduced by several orders of magnitude.

---

*Supported in part by the Fonds National Suisse.

As in the two dimensional case second order analytic results can be obtained which will allow for further improvements, and on the bases of related recent analytic results [21] we expect that analogous artificial boundary conditions will be obtained for Reynolds numbers beyond the critical point where stationary solutions become unstable and give rise to stable time periodic solutions. Finally, for the case of particle sedimentation, one also has to be able to handle finite Rossby numbers since non-symmetric free falling particles typically rotate. See [10] for a recent discussion of this problem.

To summarize, the purpose of this paper is: first, to provide a simple tool that allows to compute efficiently stationary exterior flows at low Reynolds numbers with high precision, and second, to bring to the reader's attention this new type of numerical schemes based on analytic work, a technique that we expect to become a standard tool for computations in the low Reynolds number regime well beyond the simple stationary case presented here.

The paper is organized as follows: In Section 2 we define the artificial boundary conditions and their dependence on the drag and lift. Section 3 is dedicated to the description of the numerical methods. In Section 4 we compare the numerical results obtained by means of the traditional constant Dirichlet boundary conditions with those obtained using the proposed adaptive boundary conditions.

## 2 Artificial Boundary Conditions

Consider a rigid body that is placed into a uniform stream of a homogeneous incompressible fluid filling up all of $\mathbf{R}^3$. This situation is modelled by the stationary Navier-Stokes equations (tildes are used to indicate dimension-full quantities)

$$-\rho(\tilde{\mathbf{u}} \cdot \nabla)\tilde{\mathbf{u}} + \mu \Delta \tilde{\mathbf{u}} - \nabla \tilde{p} = 0 , \qquad \nabla \cdot \tilde{\mathbf{u}} = 0 , \tag{1}$$

in $\tilde{\Omega} = \mathbf{R}^3 \setminus \tilde{\mathbf{B}}$, subject to the boundary conditions $\tilde{\mathbf{u}}|_{\partial \tilde{\mathbf{B}}} = 0$ and $\lim_{|\tilde{\mathbf{x}}| \to \infty} \tilde{\mathbf{u}}(\tilde{\mathbf{x}}) = \tilde{\mathbf{u}}_\infty$. Here, the body $\tilde{\mathbf{B}}$ is a compact set of diameter $A$ containing the origin of our coordinate system, $\tilde{\mathbf{u}}$ is the velocity field, $\tilde{p}$ is the pressure and $\tilde{\mathbf{u}}_\infty$ is some constant nonzero vector field which we choose without restriction of generality to be parallel to the $\tilde{x}$-axis, i.e., $\tilde{\mathbf{u}}_\infty = u_\infty \mathbf{e}_1$, where $\mathbf{e}_1 = (1,0,0)$ and $u_\infty > 0$. The density $\rho$ and the viscosity $\mu$, are arbitrary positive constants. From $\mu$, $\rho$ and $u_\infty$ we can form the length $\ell = \mu/(\rho u_\infty)$, the so called viscous length of the problem. The viscous forces and the inertial forces are quantities of comparable size if the Reynolds number $\mathrm{Re} = A/\ell$ is neither too small nor very large.

Below, when solving the problem (1) numerically for the example case where $\tilde{\mathbf{B}}$ is a prism, we restrict the equations (1) from the exterior infinite domain $\tilde{\Omega}$ to a sequence of bounded domains $\tilde{\mathbf{D}} \subset \tilde{\Omega}$ and study the precision of the results as a function of the domain size, once with naïve boundary conditions on the surface $\tilde{\Gamma} = \partial \tilde{\mathbf{D}} \setminus \partial \tilde{\mathbf{B}}$ of the truncated domain and once with the newly proposed adaptive boundary conditions. Note that, in contrast to the finite volume case, the boundary conditions at infinity do not prescribe the total flux of fluid (from left to right say). In particular, it does not follow from $\lim_{|\mathbf{x}|\to\infty}(\tilde{\mathbf{u}}(\mathbf{x}) - \tilde{\mathbf{u}}_\infty) = 0$, that $\lim_{x\to\infty} \int_{\mathbf{R}^2} (\tilde{\mathbf{u}} - \tilde{\mathbf{u}}_\infty)(x, \mathbf{y}) \cdot \mathbf{e}_1 \, d\mathbf{y} = 0$, the correct and non-zero value of such integrals being intimately related to the forces that act on the body (see below). It is this fact which makes the numerical implementation of (1) challenging, since any prescription of a vector field on the boundary $\tilde{\Gamma}$ also fixes the total flux across the finite domain $\tilde{\mathbf{D}}$, and it is therefore in a way quite astonishing that the solutions of (1) converge, for an increasing sequence of domains $\tilde{\mathbf{D}}$ with constant Dirichlet boundary conditions on $\tilde{\Gamma}$, to the solution of the infinite volume problem. This convergence is however very slow, since an artificial back-flow of small amplitude has to be created on a big portion of the domain $\tilde{\mathbf{D}}$ in order to accommodate for the zero flux condition for $\tilde{\mathbf{u}} - \tilde{\mathbf{u}}_\infty$ enforced by constant Dirichlet boundary conditions (see upper half of Figure 1). With our adaptive boundary conditions the flux through $\tilde{\mathbf{D}}$ is "exactly right", and no portion of the volume $\tilde{\mathbf{D}}$ is lost for the computation of a non-physical back-flow: the fluid is transported within the wake towards the body, and this fluid is then "radiated" away from the body by a source-like contribution in the velocity field $\tilde{\mathbf{u}} - \tilde{\mathbf{u}}_\infty$ (see lower half of Figure 1).

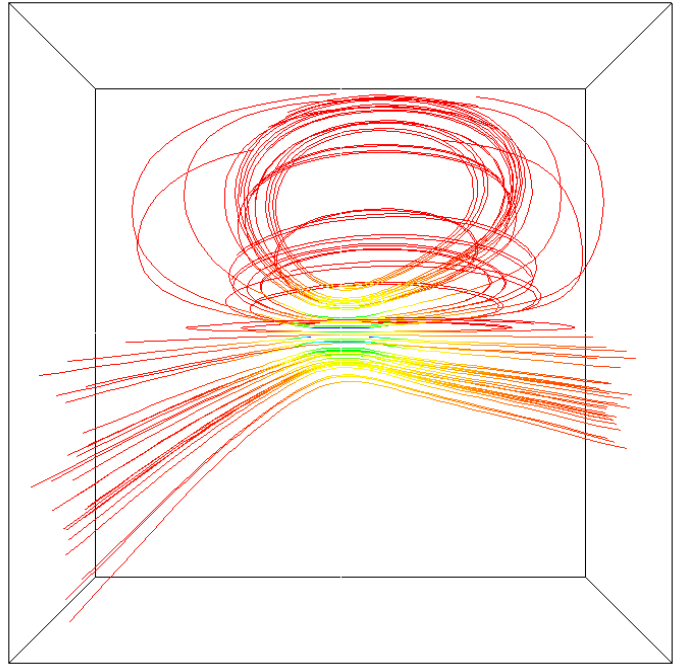

Figure 1. Streamlines, upper half: the typical nonphysical back-flow in the velocity field $\tilde{\mathbf{u}} - \tilde{\mathbf{u}}_\infty$ when imposing constant Dirichlet boundary conditions. Streamlines, lower half: no back-flow is created with adaptive boundary conditions.

For the purpose of specifying the boundary conditions we now rewrite the Navier-Stokes equations in dimensionless form. Namely, we define dimensionless coordinates $\mathbf{x} = \tilde{\mathbf{x}}/\ell$, and introduce a dimensionless vector field $\mathbf{u}$ and a dimensionless pressure $p$ through the equations $\tilde{\mathbf{u}}(\tilde{\mathbf{x}}) = u_\infty \mathbf{u}(\mathbf{x})$, $\tilde{p}(\tilde{\mathbf{x}}) = \left(\rho u_\infty^2\right) p(\mathbf{x})$. In the new coordinates we get instead of (1) the equations

$$-(\mathbf{u} \cdot \nabla)\mathbf{u} + \Delta \mathbf{u} - \nabla p = 0 \;, \qquad \nabla \cdot \mathbf{u} = 0 \;, \tag{2}$$

in $\Omega = \mathbf{R}^3 \setminus \mathbf{B}$, with the boundary conditions $\mathbf{u}|_{\partial \mathbf{B}} = 0$ and $\lim_{|\mathbf{x}| \to \infty} \mathbf{u}(\mathbf{x}) = \mathbf{e}_1$, and $\mathbf{B}$ is the set of points $\mathbf{x} \in \mathbf{R}^3$ such that $\ell \mathbf{x} \in \tilde{\mathbf{B}}$. The diameter of $\mathbf{B}$ is $A/\ell = \mathrm{Re}$ (Reynolds number). In (2) all the derivatives are with respect to the new coordinates. In what follows we use the notation $\mathbf{x} = (x, \mathbf{y})$ with $\mathbf{y} = (y_1, y_2)$ and $\mathbf{u} = (u, \mathbf{v})$ with $\mathbf{v} = (v_1, v_2)$ and we furthermore set $r = |\mathbf{x}| = (x^2 + y^2)^{1/2}$, where $y = |\mathbf{y}| = (y_1^2 + y_2^2)^{1/2}$. For Re small enough and domains $\mathbf{B}$ with smooth boundaries, equation (2) is known to have a unique classical solution [11]. In [27] this solution has been studied in detail in the half-space $\mathbf{D}_+ = \left\{(x, \mathbf{y}) \in \mathbf{R}^3 \,|\, x \geq x_+ \gg 1\right\}$, and an explicit expression for the downstream asymptotic behavior has been obtained. Global results, but in less explicit form, can also be found in [27]. Using the same ideas as in [5] one can reconstruct from the down-stream behavior of the solution the asymptotics in any direction far away from the body. Explicitly one obtains (in dimension-full variables) that, far away from the obstacle $\tilde{\mathbf{u}} \approx \tilde{\mathbf{u}}_{ABC}$, where

$$\tilde{\mathbf{u}}_{ABC}(\tilde{\mathbf{x}}) = u_\infty \mathbf{u}_{ABC}\!\left(\frac{\tilde{\mathbf{x}}}{\ell}\right) , \tag{3}$$

and where (in dimension-less variables)

$$\mathbf{u}_{ABC} = (u_{ABC}, \mathbf{v}_{ABC}) \;, \tag{4}$$
$$\mathbf{v}_{ABC} = \mathbf{v}_{1,ABC} + \mathbf{v}_{2,ABC} \;, \tag{5}$$

with

$$u_{ABC}(x,\mathbf{y}) = 1 + \frac{\theta(x)}{4\pi x}e^{-\frac{y^2}{4x}}\,c + \frac{1}{2\pi}\frac{x}{r^3}\,d + \frac{1}{2\pi}\frac{\mathbf{y}\cdot\mathbf{b}}{r^3}\,,\tag{6}$$

$$\mathbf{v}_{1,ABC}(x,\mathbf{y}) = \frac{\mathbf{y}}{8\pi x^2}\theta(x)e^{-\frac{y^2}{4x}}\,c + \frac{1}{2\pi}\frac{\mathbf{y}}{r^3}\,d$$
$$-\frac{1}{2\pi}\frac{1}{r}\frac{\text{sign}(x)}{r+|x|}\left(\mathbf{b} - \frac{1}{r}\left(\frac{1}{r} + \frac{1}{r+|x|}\right)(\mathbf{y}\cdot\mathbf{b})\,\mathbf{y}\right)\,,\tag{7}$$

$$\mathbf{v}_{2,ABC}(x,\mathbf{y}) = -\frac{\theta(x)}{2\pi x}\left(e^{-\frac{y^2}{4x}} + \frac{1}{2}\frac{e^{-\frac{y^2}{4x}}-1}{\frac{y^2}{4x}}\right)\mathbf{b}$$
$$+\frac{\theta(x)}{2\pi x}\left(\frac{e^{-\frac{y^2}{4x}}-1}{\frac{y^2}{4x}} + e^{-\frac{y^2}{4x}}\right)\frac{(\mathbf{y}\cdot\mathbf{b})}{y^2}\mathbf{y}\,,\tag{8}$$

with $c = -2d$, with $\theta$ the Heaviside function (i.e., $\theta(x) = 1$ for $x > 0$ and $\theta(x) = 0$ for $x \leq 0$), with $\text{sign}(x) = -1 + 2\theta(x)$, and with $\mathbf{y}\cdot\mathbf{b} = y_1 b_1 + y_2 b_2$. Note that the vector fields $(u_{ABC}, \mathbf{v}_{1,ABC})$ and $(0, \mathbf{v}_{2,ABC})$ are divergence free.

It is the vector field $\tilde{\mathbf{u}}_{ABC}$ that we propose to use in order to prescribe artificial boundary conditions on $\tilde{\Gamma}$. For the (positive) number $d$ and the vector $\mathbf{b}$ one has

$$d = \frac{1}{2\rho\ell^2 u_\infty^2}\tilde{F}\,,\tag{9}$$

$$\mathbf{b} = \frac{1}{2\rho\ell^2 u_\infty^2}\tilde{\mathbf{L}}\,,\tag{10}$$

where $\tilde{\mathbf{F}} = (\tilde{F}, \tilde{\mathbf{L}})$ is the force acting on the body, i.e., $\tilde{F}$ is the drag and $\tilde{\mathbf{L}}$ the lift (dimension-full quantities). The relations (9) and (10) can be obtained by using the integral form of the Navier-Stokes equations. Namely, let $D(s)$ be a disk in the $y_1, y_2$ plane of radius $s$, then we we integrate for $r$ and $s$ large enough the Navier-Stokes equations over the domain $S = [-r, r] \times D(s) \setminus \tilde{\mathbf{B}}$ and apply Gauss' theorem, which leads to a relation between the integral of the stress tensor over the body surface $\partial\tilde{\mathbf{B}}$, which gives $\tilde{\mathbf{F}}$, and the integral of the same quantity over the surface $\partial S \setminus \partial\tilde{\mathbf{B}}$. The latter integral can be computed by using the above asymptotic expressions for the velocity field and the asymptotic expression $\tilde{p} \approx -\frac{1}{2}(|\tilde{\mathbf{u}}_{ABC}|^2 - 1)$ for the pressure, in the limit where first $s \to \infty$ and then $r \to \infty$.

## 3 Solution process

In what follows we describe the discretization method and the solution process considered to solve numerically (1) on a bounded domain by means of the artificial boundary conditions. To unburden the notation we suppress throughout this section the "tildes".

### 3.1 Galerkin finite element discretization

In order to solve equation (1), we consider a discretization based on conforming mixed finite elements with continuous pressure. This discretization starts from a variational formulation of the system of equations (1) on a bounded domain $\mathbf{D} \subset \mathbf{R}^3$ containing the body $\mathbf{B}$. First, we introduce some notation needed for the derivation of this formulation.

For a bounded domain $\mathbf{D} \subset \mathbf{R}^3$, let $L^2(\mathbf{D})$ denote the Lebesgue space of square-integrable functions on $\mathbf{D}$ equipped with the inner product and norm

$$(f, g)_\mathbf{D} = \int_\mathbf{D} fg\,d\mathbf{x}\,,\qquad ||f||_\mathbf{D} = (f, f)_\mathbf{D}^{1/2}\,.$$

The pressure is assumed to lie in the space $L_0^2(\mathbf{D}) := \{q \in L^2(\mathbf{D}) \mid \int_\mathbf{D} q\,d\mathbf{x} = 0\}$, which defines it uniquely. The $L^2(\mathbf{D})$ functions with generalized (in the sense of distributions) first-order derivatives in $L^2(\mathbf{D})$ form the Sobolev space $H^1(\mathbf{D})$, while $H_0^1(\mathbf{D}) := \{v \in H^1(\mathbf{D}) \mid v|_{\partial\mathbf{D}} = 0\}$. Let $W = [H_0^1(\mathbf{D})]^3 \times L_0^2(\mathbf{D})$. For $\mathbf{w} = \{\mathbf{v}, p\} \in W$ and $\phi = \{\varphi, q\} \in W$, we define the semi-linear form

$$\mathcal{A}(\mathbf{w}; \phi) = \rho\left(((\mathbf{v} + \mathbf{u}_\infty)\cdot\nabla)\mathbf{v}, \varphi\right)_\mathbf{D} - (p, \nabla\cdot\varphi)_\mathbf{D} + 2\mu\int_\mathbf{D}\mathcal{D}(\mathbf{v}):\mathcal{D}(\varphi)\,d\mathbf{x} - (\nabla\cdot\mathbf{v}, q)_\mathbf{D}\,,\tag{11}$$

which is obtained by testing the equations (1) with $\phi \in W$ and by integration by parts of the diffusive term and the pressure gradient. $\mathcal{D}(\mathbf{v})$ denotes the deformation tensor, *i.e.*, $\mathcal{D}(\mathbf{v}) = \frac{1}{2}(\nabla \mathbf{v} + (\nabla \mathbf{v})^T)$. Then, a weak form of the equations (1) can be formulated as: find $\mathbf{w} = \{\mathbf{v}, p\} \in W$, such that

$$\mathcal{A}(\mathbf{w}; \phi) = 0 , \quad \forall \phi \in W . \tag{12}$$

The discretization of problem (12) uses a conforming finite element space $W_h \subset W$ defined on quasi-uniform triangulations $\mathcal{T}_h = \{K\}$ consisting of quadrilateral cells $K$ covering the domain $\mathbf{D}$. We consider the standard Hood-Taylor finite elements [16] for the trial and test spaces, *i.e.*, we define

$$W_h = \left\{ (\mathbf{v}, p) \in [C(\overline{\mathbf{D}})]^4 \mid \mathbf{v}|_K \in [Q_2]^3, \ p|_K \in Q_1 \right\} ,$$

where $Q_r$ describes the space of iso-parametric tensor-product polynomials of degree $r$ (for a detailed description of this standard construction process see for example [4]). This choice for the trial and test functions guarantees a stable approximation of the pressure since the Babuska-Brezzi inf-sup stability condition is satisfied uniformly in $\mathbf{D}$. The advantage, when compared to equal order function spaces for the pressure and the velocity, is that no additional pressure stabilization terms are needed. The discrete counterpart of problem (12) then reads: find $\mathbf{w}_h = \{\mathbf{v}_h, p_h\} \in \mathbf{w}_{b,h} + W_h$, such that

$$\mathcal{A}(\mathbf{w}_h; \phi_h) = 0 , \quad \forall \phi_h \in W_h . \tag{13}$$

Here $\mathbf{w}_{b,h}$ describes the prescribed Dirichlet data on the boundary $\mathbf{\Gamma}$ of the domain $\mathbf{D}$. It has to be noted that the discretization error resulting from the discrete formulation (13) involve two contributions: the discretization error due to the finite element discretization and the error contribution resulting from the approximation of $\mathbf{R}^3$ by means of the bounded domain $\mathbf{D}$. Clearly the considered domain $\mathbf{D}$ should be chosen large enough such that the associated error contribution is smaller than the error contribution due to the finite element approximation. The problem of equilibrating the error contribution during the computation can be solved by means of an adequate a posteriori error estimator. In that context a straightforward approach consists in imposing constant Dirichlet boundary conditions for $\mathbf{v}$ on $\Gamma$ which however leads to extremely large and intractable discrete problems (see [5, 6]). Our goal is to avoid these difficulties by imposing adequate non-constant Dirichlet boundary conditions on $\mathbf{\Gamma}$. In this paper we do not address the issue of deriving a posteriori error estimator in order to equilibrate the two error contributions. For more details with respect to this issue we refer to [15]. Our goal is rather to validate the proposed artificial boundary conditions and quantify the impact of these boundary conditions on the accuracy of the solution assuming the finite element contribution to be smaller.

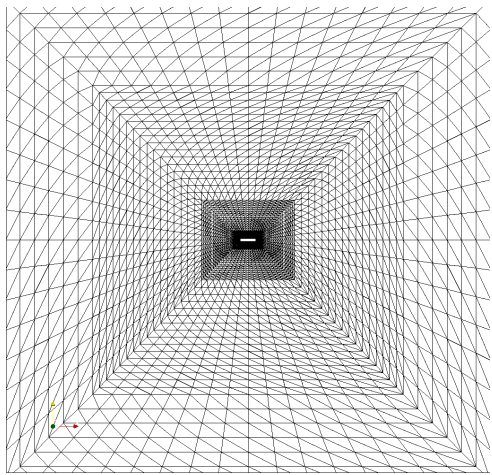

Figure 2. A typical grid used for the numerical solution of the test problem described in Section 4.

### 3.2 Computation of the drag

As explained in Section 2, the proposed artificial boundary conditions are independent of the details of the geometry of the body but depend explicitly on drag and lift. The accurate determination of these forces is therefore a key issue in our context. As in [5, 6] we use the approach proposed in [12] which is based on a reformulation of the expressions for drag and lift in terms of volume integrals by means of integration by part *i.e.* instead of considering the classical formulation

$$N_\psi(w) = \int_{\partial \mathbf{B}} (\sigma(v,p) \cdot n) \cdot \psi \, , \tag{14}$$

which represents the force acting on the rigid body $\mathbf{B}$ in the direction $\psi$ on $\partial \mathbf{B}$ one consider the equivalent formulation

$$N_\psi(w) = \mathcal{A}(w, \tilde{\psi}) \, , \tag{15}$$

where $\tilde{\psi} \in H^1(D)$ defined such that

$$\tilde{\psi} = \psi \text{ on } \partial \mathbf{B} \, , \qquad \tilde{\psi} = 0 \text{ on } \Gamma \, . \tag{16}$$

We define the discrete counterpart $N_\psi^h(w)$ of $N_\psi(w)$ by the equation

$$N_\psi^h(w_h) = \mathcal{A}(w_h, \tilde{\psi}) \, . \tag{17}$$

It is important to note that $N_\psi^h(w_h) \neq N_\psi(w_h)$. The reformulation with $N_\psi^h(w_h)$ allows to attain the full order of convergence for the values of drag and lift which can be shown to be $O(h^4)$ in our context (see [12] for more details). Convergence records of the drag for the symmetric case are plotted in Table 1, and of the drag and the lift for the non-symmetric case are plotted in Table 2. See Section 4 for details concerning the test cases.

| Level | # Unknowns | drag computed by means of | | | |
|---|---|---|---|---|---|
| | | $N_\psi(w_h)$ | | $N_\psi^h(w_h)$ | |
| 3 | 389,560 | | 5.20720 | | 4.77064 |
| 4 | 3,093,096 | | 5.16267 | | 4.75981 |
| 5 | **24,658,120** | | 5.04070 | | 4.75453 |
| Extrapolated value | | $O(h^2)$ | 5.00000 | $O(h^4)$ | 4.75418 |

Table 1. Convergence records of the drag for the symmetric case at Re = 1 considering both formulations $N_\psi^h(\mathbf{w}_h)$ and $N_\psi(\mathbf{w}_h)$ using adaptive boundary conditions. The computational domain has a diameter that is hundred times bigger than that of the rigid body ($\text{diam}(D) = 100$). Note that due to the symmetry of the test problem the lift is in this configuration equal to zero. With more than twenty million unknowns Level 5 involves large computations which need to be executed on a HPC platform. The resulting value of the drag is after extrapolation used as the reference value for the computations of relative errors when computing the drag on smaller domains and for other boundary conditions.

| Level | # Unknowns | drag computed by means of | | | | lift computed by means of | | | |
|---|---|---|---|---|---|---|---|---|---|
| | | $N_\psi(w_h)$ | | $N_\psi^h(w_h)$ | | $N_\psi(w_h)$ | | $N_\psi^h(w_h)$ | |
| 3 | 389,560 | | 5.22018 | | 4.78124 | | -0.15182 | | -0.12436 |
| 4 | 3,093,096 | | 5.17497 | | 4.77037 | | -0.14401 | | -0.12387 |
| 5 | **24,658,120** | | 5.05246 | | 4.76507 | | -0.13782 | | -0.12365 |
| Extrapolated value | | $O(h^2)$ | 5.0116 | $O(h^4)$ | 4.7647 | $O(h^2)$ | -0.13576 | $O(h^4)$ | -0.12364 |

Table 2. Convergence records of the drag and the lift for the non-symmetric case at Re = 1 considering both formulations $N_\psi^h(\mathbf{w}_h)$ and $N_\psi(\mathbf{w}_h)$ using adaptive boundary conditions. The computational domain has a diameter that is hundred times bigger than that of the rigid body ($\text{diam}(D) = 100$). With more than twenty million unknowns Level 5 involves large computations which need to be executed on a HPC platform. The resulting values of drag and lift are after extrapolation used as reference values for the computation of relative errors when computing drag and lift on smaller domains and for other boundary conditions.

### 3.3 Solver

The specificity of the proposed approach is related to the fact that the prescribed boundary conditions depend on the drag and lift acting on the rigid body **B**. There is therefore a coupling between the definition of the boundary condition and the solution $\mathbf{w}_h = \{\mathbf{v}_h, p_h\}$ of (13) to be computed. Numerical experience show that this coupling can be treated by means of a fixed iteration where the boundary conditions are determined by means of successive updates, based on previously computed values of the drag and lift (see Algorithm 1).

The nonlinear algebraic system (13) is solved implicitly in a fully coupled manner by means of a damped Newton method. Denoting the derivative of $\mathcal{A}(\cdot, \cdot)$ taken at a discrete function $\mathbf{w}_h \in W_h$ by $\mathcal{A}'(\mathbf{w}_h, \cdot)(\cdot)$, the linear system arising at the Newton step number $k$ has the following form,

$$\mathcal{A}'(\mathbf{w}_h^k, \phi_h)(\hat{\mathbf{w}}_h^k) = (\mathbf{r}_h^k, \phi_h) , \qquad \forall \phi_h \in W_h , \tag{18}$$

where $\mathbf{r}_h^k$ is the equation residual of the current approximation $\mathbf{w}_h^k$, and where $\hat{\mathbf{w}}_h^k$ corresponds to the needed correction. The updates $\mathbf{w}_h^{k+1} = \mathbf{w}_h^k + \alpha^k \hat{\mathbf{w}}_h^k$ with a relaxation parameter $\alpha^k$ chosen by means of Armijo's rule are carried out until convergence. In practice, the Jacobian involved in (18) is directly derived from the analytical expression for the derivative of the variational system (13).

It is well known that the ability of the Newton iteration to converge at the local rate greatly depends on the quality of the initial approximation (see *e.g.* [17]). In order to find such an initial approximation, we consider a mesh hierarchy $\mathcal{T}_{h_l}$ with $\mathcal{T}_{h_l} \subset \mathcal{T}_{h_{l+1}}$, and the corresponding system of equations (13) is successively solved by taking advantage of the previously computed solution, *i.e.*, the nonlinear Newton steps are embedded in a nested iteration process (see *e.g.* [24], chapter 8).

The linear subproblems (18) are solved by the *Generalized Minimal Residual Method* (GMRES), see [20], preconditioned by means of multigrid iterations. See [24] and references therein for a description of the different multigrid techniques for flow simulations. This preconditioner, based on a new multigrid scheme oriented towards conformal higher order finite element methods, is a key ingredient of the overall solution process. Two specific features characterizing the proposed scheme are: varying order of the finite element Ansatz on the mesh hierarchy and a Vanka type smoother [22] adapted to higher order discretization. This somewhat technical part of the solver is described in full details in [14].

---

**Algorithm 1** Overall solution process related to the artificial boundary conditions

---

$d_0 = 0$, $c_0 = 0$ (corresponds to homogeneous Dirichlet boundary condition)
**for** $i = 0, ...$ **do**
  **i) Solve** the discrete Navier-Stokes problem (13) assuming $d = d_i$ and $b = b_i$ for the boundary conditions leading to the solution $\mathbf{w}_h^i = \{\mathbf{v}_h^i, p_h^i\}$
  **ii) Compute** $D_r = Drag(\mathbf{w}_h^i)$ and $L_f = Lift(\mathbf{w}_h^i)$
  **iii) If** convergence attained for $D_r$ and $L_f$ **then** STOP
  **iv) Update** boundary conditions i.e. $d_{i+1} = d(D_r, L_f)$ and $b_{i+1} = b(D_r, L_f)$

---

## 4 Numerical experiments

The goal of this section is twofold. First we compare, for a given configuration of the flow, the solution obtained by means of the proposed adaptive boundary conditions with the solution obtained by means of constant Dirichlet boundary conditions. The results confirm the expectations based on the theoretical results and provide numerical evidence for the validity of our approach. Second we provide quantitative results for different Reynolds numbers, which clearly show the drastic improvement in numerical efficiency that results from the proposed boundary conditions.

### 4.1 Description of the configuration

Our model problem consists of a prism $[-0.5, 0.5] \times [-0.05, 0.05]^2$ which is immersed into a uniform stream of a homogeneous incompressible fluid with density $\rho = 1$ and dynamic viscosity $\mu = 1$. Further we impose $u_\infty = 1$. With $A = 1$ being the length of the prism, we find for the Reynolds number corresponding to this configuration Re = 1. For the computational domain **D** we use spheres centered at the origin. Since the prism is aligned with the $x$-axis, and therefore with the flow at infinity, the resulting flow is symmetric with respect to the $x$-$z$ plane and the $x$-$y$ plane and no lift is produced. We therefore also consider a second configuration where we tilt the prism by 10 degrees about the $y$-axis. The flow is still symmetric with respect to the $x$-$z$ plane but non-symmetric with respect to the $x$-$y$ plane, and lift is produced in the negative $z$ direction.

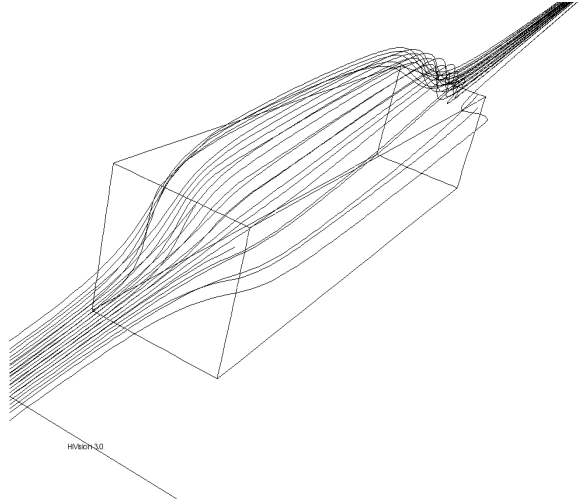

Figure 3. Streamlines around the body for the non-symmetric test case at Re = 1.

## 4.2 Convergence along cuts

We now discuss various ways of comparing the numerical solutions $\mathbf{u} = (u, \mathbf{v})$ of (1) obtained with constant Dirichlet and with the adaptive boundary conditions with the theoretical predictions. For this purpose we define so called "cuts", *i.e.*, we plot components of the vector field along certain lines for which we have precise theoretical predictions. These cuts and the corresponding theoretical predictions are defined in the following table:

| Notation | Definition | Scaled Velocity | Asymptotic |
|---|---|---|---|
| **cutI**$_u$ | $x > 0, y = 0, z = 0$ | $-2\pi x u$ | $d$ for $x \to \infty$ |
| **cutI**$_{v_2}$ | $x > 0, y = 0, z = 0$ | $-4\pi x v_2$ | $b_2$ for $x \to \infty$ |
| **cutIIa**$_{v_1}$ | $x = 0, y > 0, z = 0$ | $2\pi y^2 v_1$ | $d - b_1$ for $y \to \infty$ |
| **cutIIa**$_{v_2}$ | $x = 0, y > 0, z = 0$ | $2\pi y^2 v_2$ | $b_2$ for $y \to \infty$ |
| **cutIIb**$_u$ | $x = 0, y = 0, z > 0$ | $2\pi z^2 u$ | $b_2$ for $z \to \infty$ |
| **cutIIb**$_{v_2}$ | $x = 0, y = 0, z > 0$ | $2\pi z^2 v_2$ | $d - b_2$ for $z \to \infty$ |
| **cutIIbm**$_u$ | $x = 0, y = 0, z < 0$ | $-2\pi z^2 u$ | $b_2$ for $z \to -\infty$ |
| **cutIIbm**$_{v_2}$ | $x = 0, y = 0, z < 0$ | $-2\pi z^2 v_2$ | $d + b_2$ for $z \to -\infty$ |
| **cutIII**$_u$ | $x < 0, y = 0, z = 0$ | $-2\pi x^2 u$ | $d$ for $x \to -\infty$ |
| **cutIII**$_{v_2}$ | $x < 0, y = 0, z = 0$ | $4\pi x^2 v_2$ | $b_2$ for $x \to -\infty$ |

Table 3. Nomenclature of the cuts considered for the numerical experiments. The notation is the same as the one used in the definition of the artificial boundary conditions (3)-(10), and the asymptotic values in the forth column are obtained by scaling the velocity components as given in (3)-(10) as indicated in Column 3 and by taking the limit specified in Column 4.

### 4.2.1 Symmetric case

The following figures summarize our numeric results for the case without lift. From the extrapolated values in Table 1 we find using the definitions (9) and (10) for the drag of the symmetric configuration $F \approx 4.754$ and therefore that $d \approx 2.377$.

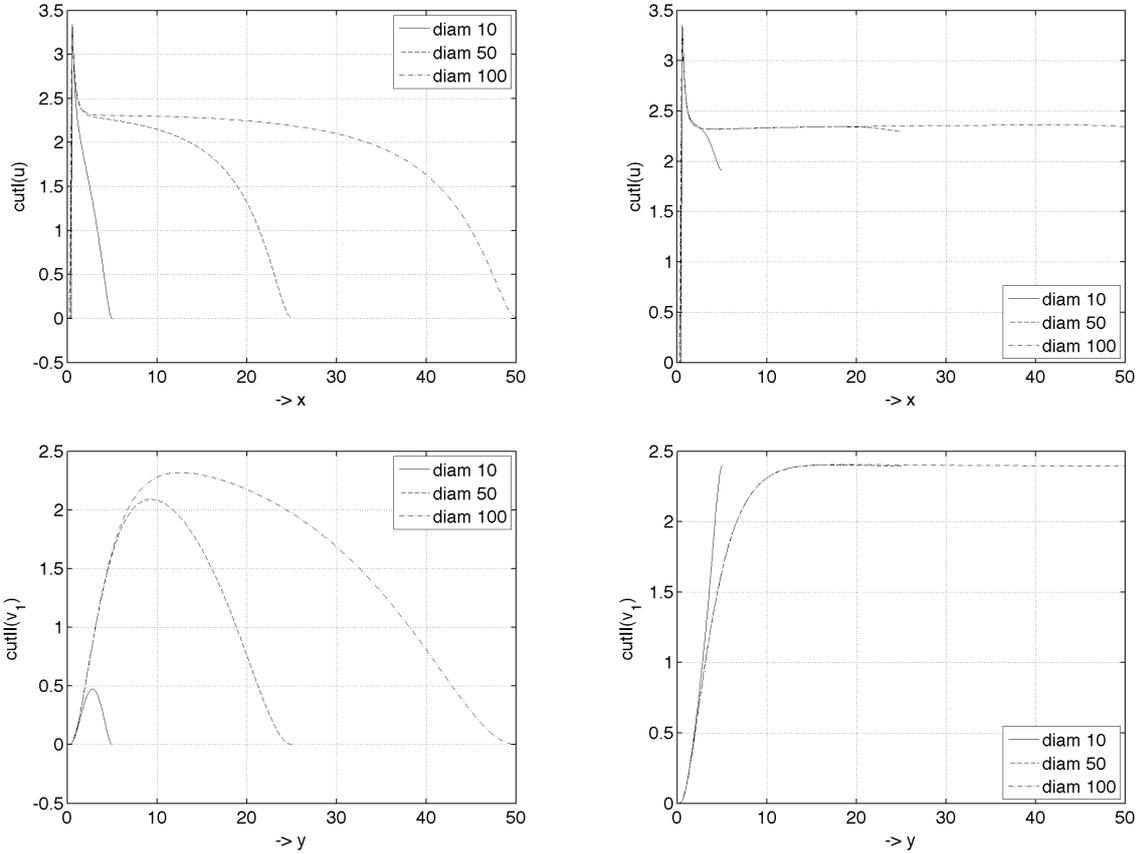

Figure 4. From above to below, scaled velocity components defined by means of $\mathbf{cutI}_u$ and $\mathbf{cutIIa}_{v_1}$ considering constant Dirichlet boundary conditions (left column) and the proposed adaptive boundary conditions (right column). The size of the computational domain $\mathbf{D}$ varies in the range $\text{diam}(\mathbf{D}) \in [10, 100]$. The considered configuration corresponds to the symmetric case at Re = 1.

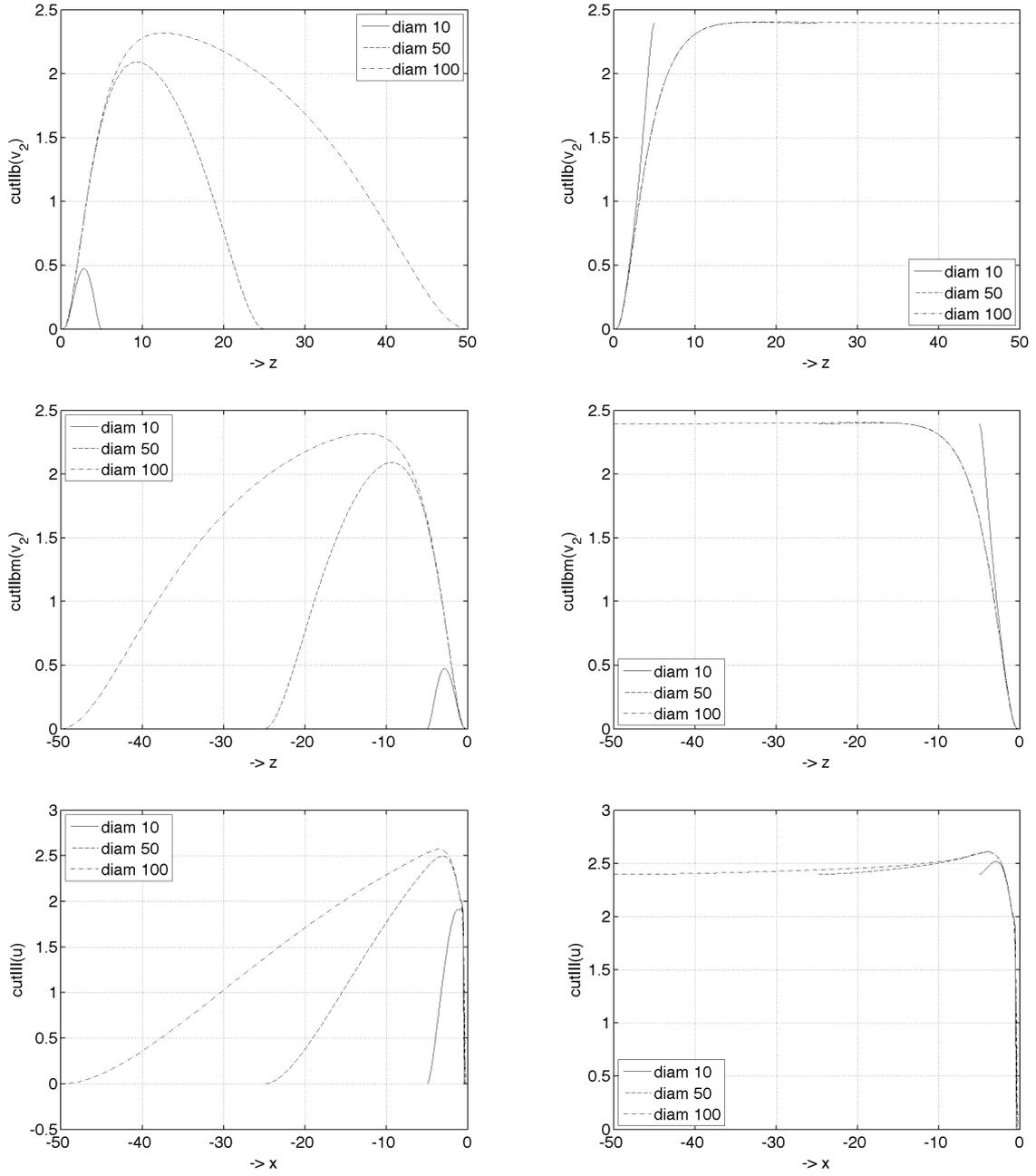

Figure 5. From above to below, scaled velocity components defined by means of **cutIIb**$_{v_2}$, **cutIIbm**$_{v_2}$, and **cutIII**$_u$ considering constant Dirichlet boundary conditions (left column) and the proposed adaptive boundary conditions (right column). The size of the computational domain **D** varies in the range diam(**D**) $\in [10, 100]$. The considered configuration corresponds to the symmetric case at Re = 1. Recall that $d \approx 2.377$.

### 4.2.2 Non-symmetric case

The following figures summarize our results for the case with lift. From the extrapolated values in Table 2 we find for the drag and lift in the $x$-$z$ plane for the non-symmetric configuration $F \approx 4.764$ and therefore that $d \approx 2.382$ and $L \approx -0.1236$ and therefore that $b_2 \approx -0.0618$. Furthermore $d - b_2 \approx 2.44$ and $d + b_2 \approx 2.32$.

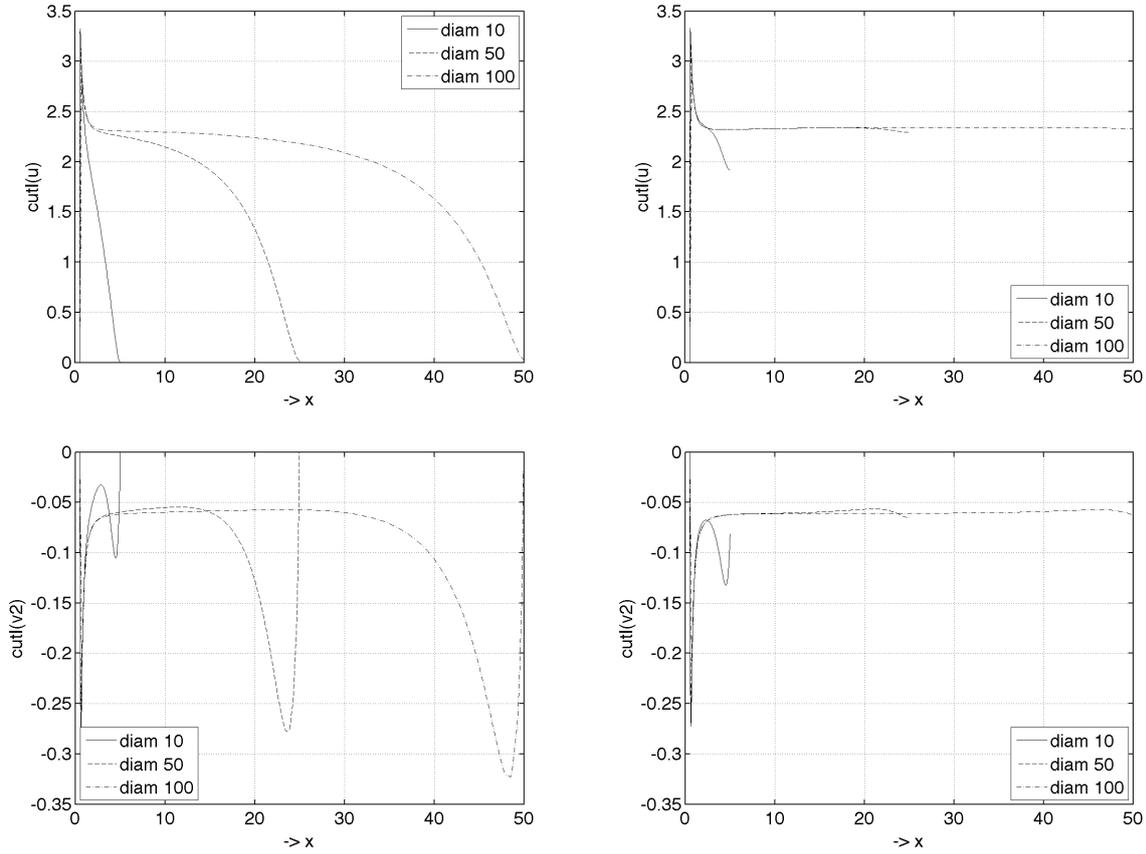

Figure 6. From above to below, scaled velocity components defined by means of **cutI**$_u$ and **cutI**$_{v_2}$ considering constant Dirichlet boundary conditions (left column) and the proposed adaptive boundary conditions (right column). The size of the computational domain **D** varies in the range diam(**D**) $\in$ [10, 100]. The considered configuration corresponds to the symmetric case at Re = 1.

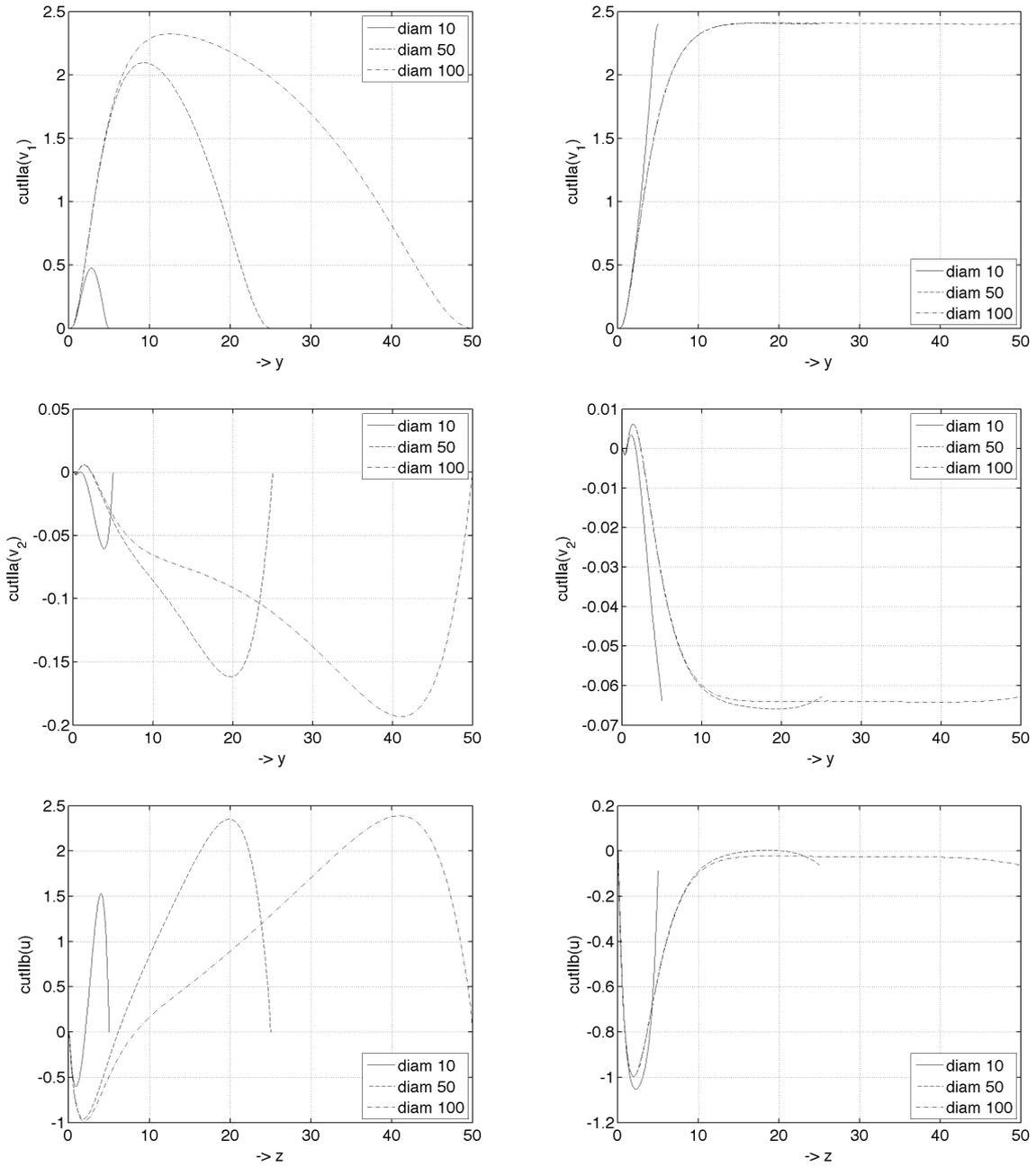

Figure 7. From above to below, scaled velocity components defined by means of **cutIIa**$_{v_1}$, **cutIIa**$_{v_2}$, and **cutIIb**$_u$ considering constant Dirichlet boundary conditions (left column) and the proposed adaptive boundary conditions (right column). The size of the computational domain **D** varies in the range $\mathrm{diam}(\mathbf{D}) \in [10, 100]$. The considered configuration corresponds to the symmetric case at Re = 1. Recall that $d \approx 2.382$ and $b_2 \approx -0.0618$.

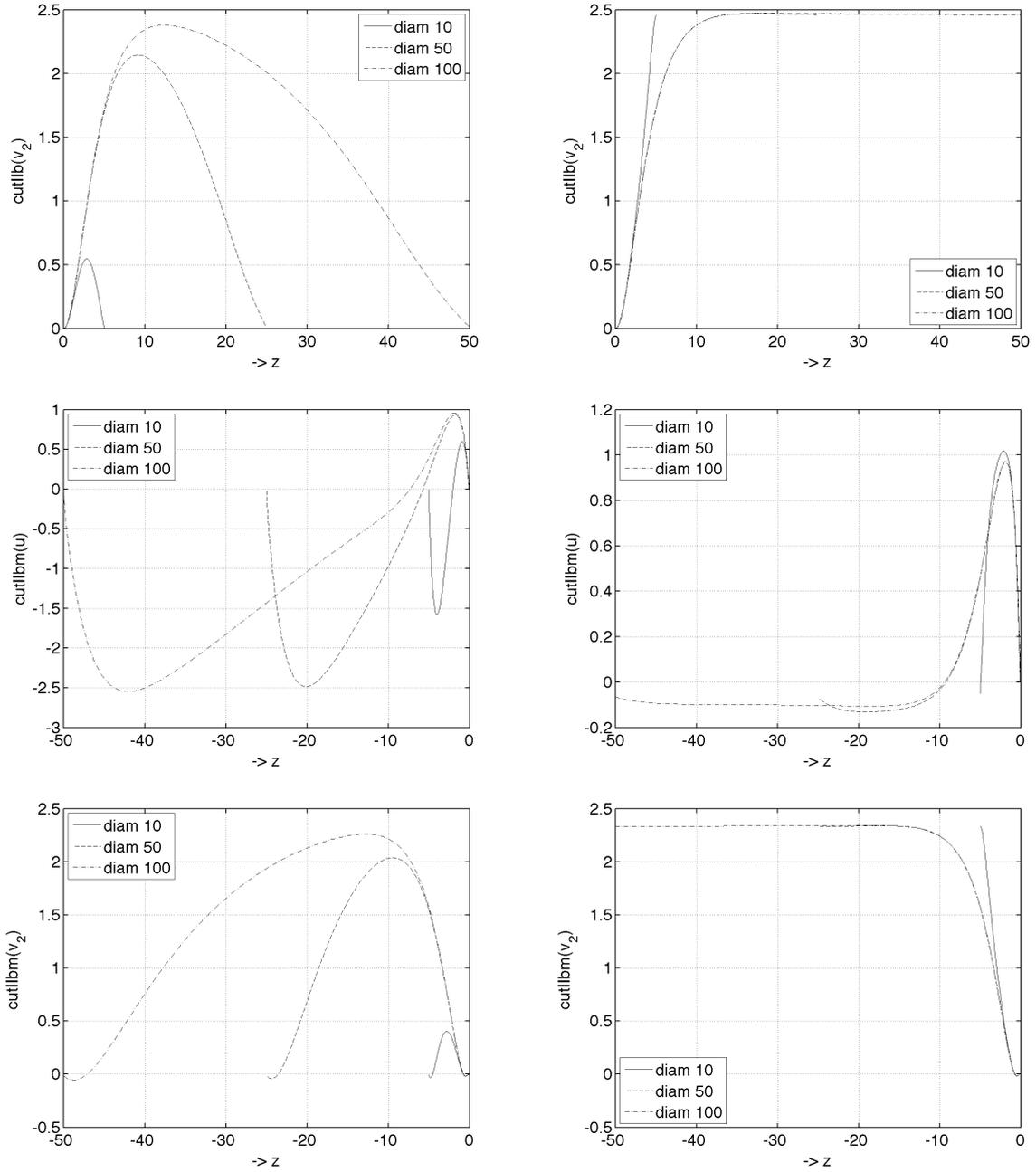

Figure 8. From above to below, scaled velocity components defined by means of **cutIIb**$_{v_2}$, **cutIIm**$_u$ and **cutIIbm**$_{v_2}$ considering constant Dirichlet boundary conditions (left column) and the proposed adaptive boundary conditions (right column). The size of the computational domain **D** varies in the range diam(**D**) ∈ [10, 100]. The considered configuration corresponds to the symmetric case at Re = 1. Recall that $b_2 \approx -0.0618$, $d - b_2 \approx 2.44$ and $d + b_2 \approx 2.32$.

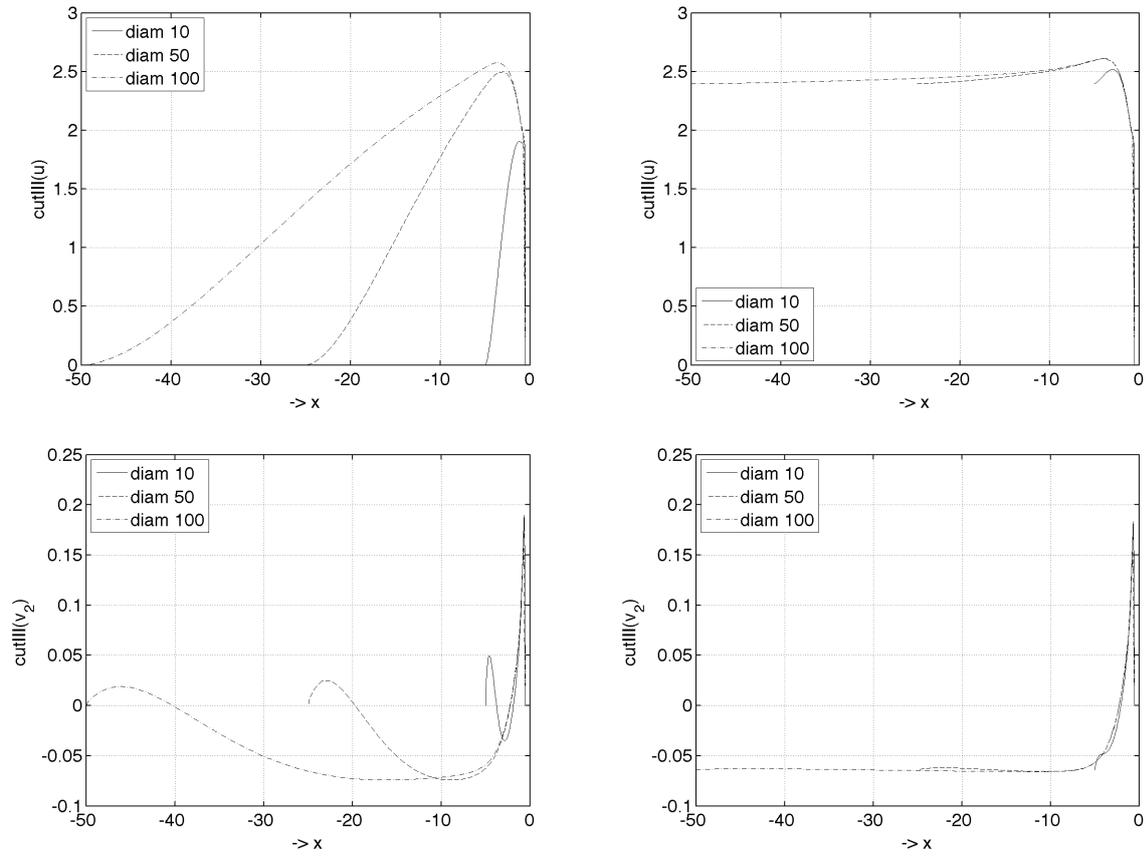

Figure 9. From above to below, scaled velocity components defined by means of **cutIII**$_u$ and **cutIII**$_{v_2}$ considering constant Dirichlet boundary conditions (left column) and the proposed adaptive boundary conditions (right column). The size of the computational domain **D** varies in the range diam(**D**) ∈ [10, 100]. The considered configuration corresponds to the symmetric case at Re = 1. Recall that $d \approx 2.382$ and $b_2 \approx -0.0618$.

## 4.3 Relative errors for drag and lift

The following figures illustrate the gain in precision for a given computational domain when using adaptive boundary conditions instead of constant Dirichlet boundary conditions. We note that the figures are the result of a very important computational effort since very precise results for the drag and lift have first to be obtained on a very large domain in order to have reference values for drag and lift with respect to which the relative errors are computed. These values are $F \approx 4.754$ for the drag in the symmetric case and $F \approx 4.764$ for the drag and $L \approx -0.1236$ for the lift for non-symmetric case and these values have been computed using the adaptive boundary conditions on a domain of diameter 100. See Table 1 and Table 2.

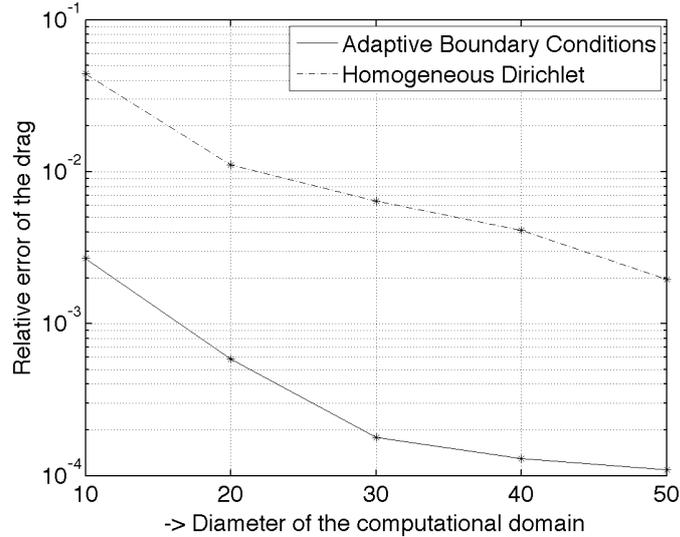

Figure 10. Plot of the relative error of the drag as a function of the domain diameter, considering constant Dirichlet boundary conditions and the adaptive boundary conditions, respectively. The plotted data are for the symmetric case at Re = 1.

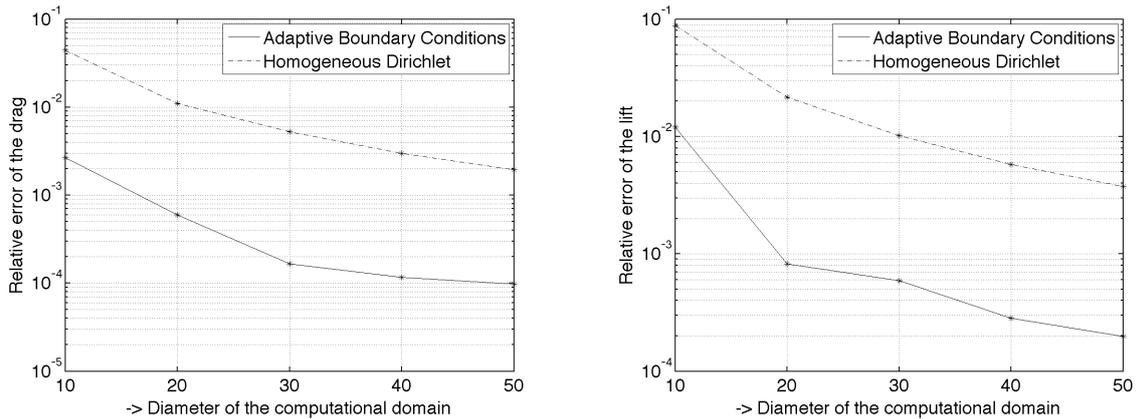

Figure 11. Plot of the relative error of the drag as a function of the domain diameter, considering constant Dirichlet boundary conditions and the adaptive boundary conditions, respectively. The plotted data are for the non-symmetric case at Re = 1.

## 5 Conclusion and outlook

We have discussed the problem of solving numerically the stationary incompressible Navier-Stokes equations in a domain exterior to a body in three dimensions. A self-consistent scheme for choosing artificial boundary conditions has been introduced, which incorporates in particular the computation of drag and lift exerted on the body as part of the solution process. When compared with the results obtained using traditional constant boundary conditions computational times are typically reduced by several orders of magnitude.

# Acknowledgement

The authors greatly thank the division "Scientific Computing and Application" from the Steinbuch Centre for Computing (SCC) at the University Karlsruhe for its excellent support needed for the extensive computation on the Supercomputer HP XC 4000 at SCC. A special thank goes to H. Gernert and R. Lohner for their kind support and many important suggestions. With respect to the code optimization on the considered parallel platform the authors greatly acknowledge H. Bockelmann for his support and strong commitment.

# References

[1] M. B. Baker. Cloud microphysics and climate. *Science*, 276:279–306, 1997.

[2] G. K. Batchelor. *An Introduction to Fluid Dynamics*. Cambridge University Press, 1967.

[3] D. Bichsel and P. Wittwer. Stationary flow past a semi-infinite flat plate: analytical and numerical evidence for a symmetry-breaking solution. *Journal of Statistical Physics*, online first, 2007.

[4] S.C. Brenner and R.L. Scott. *The mathematical theory of finite element methods*. Springer, Berlin-Heidelberg-New York, 1994.

[5] S. Bönisch, V. Heuveline, and P. Wittwer. Adabtive boundary conditions for exterior flow problems. *Journal of Mathematical Fluid Mechanics*, 7:85–107, 2005.

[6] S. Bönisch, V. Heuveline, and P. Wittwer. Second order adaptive boundary conditions for exterior flow problems: non-symmetric stationary flows in two dimensions. *Journal of mathematical fluid fechanics*, 8:1–26, 2006.

[7] B. H. Carmichael. Low Reynolds number airfoil survey. Technical report, NASA, 1981.

[8] R. Eppler and D. M. Somers. Low speed airfoil design and analysis. *Advanced Technology Airfoil Research, Volume I*, 1:73–100, 1979.

[9] R. Eppler and D. M. Somers. *A Computer Program for the Design and Analysis of Low-Speed Airfoils*, 1981.

[10] G. P. Galdi and A. L. Silvestre. Strong solutions of the Navier-Stokes equations around a roating obstacle. *Arch. Rational Mech. Anal.*, 176(3):331–350, 2005.

[11] G.P. Galdi. *An introduction to the mathematical theory of the Navier-Stokes equations: Nonlinear steady problems*. Springer Tracts in Natural Philosophy, Vol. 39, Springer-Verlag, 1998.

[12] M. Giles, M. Larson, M. Levenstam, and E. Süli. Adaptive error control for finite element approximations of the lift and drag coefficients in viscous flow. Technical Report NA-97/06, Oxford University Computing Laboratory, 1997.

[13] F. Haldi and P. Wittwer. leading order down-stream asymptotics of non-symmetric stationary Navier-Stokes flows in two dimensions. *journal of mathematical fluid mechanics*, 7:611–648, 2005.

[14] V. Heuveline. On higher-order mixed FEM for low Mach number flows: Application to a natural convection benchmark problem. *Int. J. Numer. Math. Fluids*, 41(12):1339–1356, 2003.

[15] V. Heuveline. Adaptive finite elements for the steady free fall of a body in newtonian fluid. *Comptes Rendus de l'Académie des Sciences, Mécanique*, 333(12):896–909, 2005.

[16] P. Hood and C. Taylor. A numerical solution of the Navier-Stokes equations using the finite element techniques. *Comp. and Fluids*, 1:73–100, 1973.

[17] C.T. Kelley. *Iterative methods for linear and nonlinear equations*, volume 16 of *Frontiers in Applied Mathematics*. SIAM, Philadelphia, 1995.

[18] J. Latt, Y. Grillet, B. Chopard, and P. Wittwer. Simulating an exterior domain for drag force computations in the lattice boltzmann method. *Math. Comp. Sim.*, 72:169–172, 2006.

[19] T. J. Mueller. *Fixed and Flapping Wing Aerodynamics for Micro Air Vehicle Applications*, volume 195. AIAA, 2001.

[20] Y. Saad. *Iterative methods for sparse linear systems*. Computer Science/Numerical Methods. PWS Publishing Company, 1996.

[21] G. van Baalen. Downstream asymptotics in exterior domains: from stationary wakes to time periodic flows. *Journal of Mathematical Fluid mechanics*, November 2006.

[22] S. Vanka. Block-implicit multigrid calculation of two-dimensional recirculating flows. *Comp. Meth. Appl. Mech. Eng.*, 59(1):29–48, 1986.

[23] P. K. Wang and W. Ji. Collision efficiencies of ice crystals at low-intermediate Reynolds numbers colliding with supercooled cloud droplets: a numerical study. *Journal of the atmospheric sciences*, 57:1001–1009, 2000.

[24] P. Wesseling. *An introduction to multigrid methods*. Wiley, Chichester, 1992.

[25] P. Wittwer. On the structure of Stationary Solutions of the Navier-Stokes equations. *Commun. Math. Phys.*, 226:455–474, 2002.

[26] P. Wittwer. Supplement: On the structure of stationary solutions of the Navier-Stokes equations. *Commun. Math. Phys.*, 234:557–565, 2003.

[27] P. Wittwer. Leading order down-stream asymptotics of stationary Navier-Stokes flows in three dimensions. *Journal of Mathematical Fluid Mechanics*, 8:147–186, 2006.